# MOA-2020-BLG-208Lb: Cool Sub-Saturn Mass Planet Within Predicted Desert

Greg Olmschenk,[1,2,3] David P. Bennett,[1,4,5,3] Ian A. Bond,[6,3] Weicheng Zang,[7,8] Youn Kil Jung,[9,10] Jennifer C. Yee,[11,10,12] Etienne Bachelet,[13,14]

(Leading authors)

Fumio Abe,[15] Richard K. Barry,[1] Aparna Bhattacharya,[1,4,5] Hirosane Fujii,[15] Akihiko Fukui,[16,17] Yuki Hirao,[18] Stela Ishitani Silva,[1,19,5] Yoshitaka Itow,[15] Rintaro Kirikawa,[18] Iona Kondo,[18] Naoki Koshimoto,[1,4,18] Yutaka Matsubara,[15] Sho Matsumoto,[18] Shota Miyazaki,[18] Brandon Munford,[20] Yasushi Muraki,[15] Arisa Okamura,[18] Clément Ranc,[21] Nicholas J. Rattenbury,[20] Yuki Satoh,[18] Takahiro Sumi,[18] Daisuke Suzuki,[18] Taiga Toda,[18] Paul J. Tristram,[22] Aikaterini Vandorou,[1,4,5] Hibiki Yama,[18]

(The MOA Collaboration)

Michael D. Albrow,[23] Sang-Mok Cha,[9,24] Sun-Ju Chung,[9,25] Andrew Gould,[26,27,12] Cheongho Han,[28] Kyu-Ha Hwang,[9] Dong-Jin Kim,[9] Hyoun-Woo Kim,[9] Seung-Lee Kim,[9,25] Chung-Uk Lee,[9] Dong-Joo Lee,[9] Yongseok Lee,[9,24] Byeong-Gon Park,[9,25] Richard W. Pogge,[29,12] Yoon-Hyun Ryu,[9] In-Gu Shin,[9] Yossi Shvartzvald,[30,8]

(The KMTNet Collaboration)

Grant Christie,[31] Tony Cooper,[32] John Drummond,[33,34] Jonathan Green,[32] Steve Hennerley,[32] Jennie McCormick,[35] L. A. G. Monard,[36] Tim Natusch,[31,37] Ian Porritt,[38] Thiam-Guan Tan,[39]

(The MicroFUN Collaboration)

Shude Mao,[7,40] Dan Maoz,[41] Matthew T. Penny,[42] Wei Zhu,[7,43]

(The MAP Follow-Up Collaboration)

V. Bozza,[44,45] Arnaud Cassan,[13] Martin Dominik,[46] Markus Hundertmark,[47] R. Figuera Jaimes,[48] K.Kruszyńska,[49] K.A.Rybicki,[49,30] R.A. Street,[50] Y. Tsapras,[47] Joachim Wambsganss,[47] Ł.Wyrzykowski,[49] P. Zieliński,[51]

(The OMEGA Collaboration)

Gioia Rau,[1,19]

—

[1] NASA Goddard Space Flight Center, Greenbelt, MD 20771, USA
[2] Oak Ridge Associated Universities, Oak Ridge, TN 37830, USA
[3] The MOA Collaboration
[4] Department of Astronomy, University of Maryland, College Park, MD 20742, USA
[5] Center for Research and Exploration in Space Science and Technology, NASA/GSFC, Greenbelt, MD 20771
[6] Institute of Natural and Mathematical Sciences, Massey University, Auckland 0745, New Zealand
[7] Department of Astronomy, Tsinghua University, Beijing 100084, China
[8] The MAP Follow-Up Collaboration
[9] Korea Astronomy and Space Science Institute, Daejon 34055, Republic of Korea
[10] The KMTNet Collaboration
[11] Center for Astrophysics | Harvard & Smithsonian, 60 Garden Street, Cambridge, MA 02138, USA
[12] The MicroFUN Collaboration
[13] Institut d'Astrophysique de Paris, Sorbonne Université, CNRS, UMR 7095, 98 bis bd Arago, 75014 Paris, France
[14] The OMEGA Collaboration
[15] Institute for Space-Earth Environmental Research, Nagoya University, Nagoya 464-8601, Japan
[16] Department of Earth and Planetary Science, Graduate School of Science, The University of Tokyo, 7-3-1 Hongo, Bunkyo-ku, Tokyo 113-0033, Japan
[17] Instituto de Astrofísica de Canarias, Vía Láctea s/n, E-38205 La Laguna, Tenerife, Spain
[18] Department of Earth and Space Science, Graduate School of Science, Osaka University, Toyonaka, Osaka 560-0043, Japan

[19]Department of Physics, The Catholic University of America, Washington, DC 20064, USA
[20]Department of Physics, University of Auckland, Private Bag 92019, Auckland, New Zealand
[21]Sorbonne Université, CNRS, Institut d'Astrophysique de Paris, IAP, F-75014, Paris, France
[22]University of Canterbury Mt. John Observatory, P.O. Box 56, Lake Tekapo 8770, New Zealand
[23]University of Canterbury, Department of Physics and Astronomy, Private Bag 4800, Christchurch 8020, New Zealand
[24]School of Space Research, Kyung Hee University, Yongin 17104, Republic of Korea
[25]University of Science and Technology, Korea, 217 Gajeong-ro Yuseong-gu, Daejeon 34113, Republic of Korea
[26]Department of Astronomy, Ohio State University, 140 W. 18th Ave., Columbus, OH 43210, USA
[27]Max-Planck-Institute for Astronomy, Königstuhl 17, D-69117 Heidelberg, Germany
[28]Department of Physics, Chungbuk National University, Cheongju 28644, Republic of Korea
[29]Department of Astronomy, The Ohio State University, 140 W. 18th Avenue, Columbus, OH 43210, USA
[30]Department of Particle Physics and Astrophysics, Weizmann Institute of Science, Rehovot 76100, Israel
[31]Auckland Observatory, Auckland, New Zealand
[32]Kumeu Observatory, Kumeu, New Zealand
[33]Possum Observatory, Patutahi, New Zealand
[34]Centre for Astrophysics, University of Southern Queensland, Toowoomba, Queensland 4350, Australia
[35]Farm Cove Observatory, Centre for Backyard Astrophysics, Pakuranga, Auckland, New Zealand
[36]Klein Karoo Observatory, Centre for Backyard Astrophysics, Calitzdorp, South Africa
[37]Institute for Radio Astronomy and Space Research (IRASR), AUT University, Auckland, New Zealand
[38]Turitea Observatory, Palmerston North, New Zealand
[39]Perth Exoplanet Survey Telescope, Perth, Australia
[40]National Astronomical Observatories, Chinese Academy of Sciences, Beijing 100101, China
[41]School of Physics and Astronomy, Tel-Aviv University, Tel-Aviv 6997801, Israel
[42]Department of Physics and Astronomy, Louisiana State University, Baton Rouge, LA 70803 USA
[43]Canadian Institute for Theoretical Astrophysics, University of Toronto, 60 St George Street, Toronto, ON M5S 3H8, Canada
[44]Dipartimento di Fisica "E.R. Caianiello", Università degli studi di Salerno, Via Giovanni Paolo II 132, I-84084 Fisciano (SA), Italy
[45]Istituto Nazionale di Fisica Nucleare, Sezione di Napoli, Via Cintia, 80126, Napoli
[46]University of St Andrews, Centre for Exoplanet Science, SUPA School of Physics & Astronomy, North Haugh, St Andrews, KY16 9SS, United Kingdom
[47]Zentrum für Astronomie der Universität Heidelberg, Astronomisches Rechen-Institut, Mönchhofstr. 12-14, 69120 Heidelberg, Germany
[48]Facultad de Ingeniería y Tecnología, Universidad San Sebastian, General Lagos 1163, Valdivia 5110693, Chile
[49]Astronomical Observatory, University of Warsaw, Al. Ujazdowskie 4, 00-478 Warszawa, Poland
[50]Las Cumbres Observatory, 6740 Cortona Drive, Suite 102, Goleta, CA 93117, USA
[51]Institute of Astronomy, Faculty of Physics, Astronomy and Informatics, Nicolaus Copernicus University in Toruń, ul. Grudziądzka 5, 87-100 Toruń, Poland

## ABSTRACT

We analyze the MOA-2020-BLG-208 gravitational microlensing event and present the discovery and characterization of a new planet, MOA-2020-BLG-208Lb, with an estimated sub-Saturn mass. With a mass-ratio $q = 3.17^{+0.28}_{-0.26} \times 10^{-4}$, the planet lies near the peak of the mass-ratio function derived by the MOA collaboration (Suzuki et al. 2016) and near the edge of expected sample sensitivity. For these estimates we provide results using two mass law priors: one assuming that all stars have an equal planet-hosting probability, and the other assuming that planets are more likely to orbit around more massive stars. In the first scenario, we estimate that the lens system is likely to be a planet of mass $m_{\rm planet} = 46^{+42}_{-24}\ M_\oplus$ and a host star of mass $M_{\rm host} = 0.43^{+0.39}_{-0.23}\ M_\odot$, located at a distance $D_L = 7.49^{+0.99}_{-1.13}$ kpc. For the second scenario, we estimate $m_{\rm planet} = 69^{+37}_{-34}\ M_\oplus$, $M_{\rm host} = 0.66^{+0.35}_{-0.32}\ M_\odot$, and $D_L = 7.81^{+0.93}_{-0.93}$ kpc. The planet has a projected separation as a fraction of the Einstein ring radius $s = 1.3807^{+0.0018}_{-0.0018}$. As a cool sub-Saturn-mass planet, this planet adds to a growing collection of evidence for revised planetary formation models.

## 1. INTRODUCTION

Gravitational microlensing (Mao & Paczynski 1991) provides a means for detecting planets that is sensitive to low-mass planets orbiting at moderate to large distances from their host star (Bennett & Rhie 1996; Gould & Loeb 1992), typically from 0.5–10 AU. Such planets may be challenging to detect via other common exoplanet detection methods (e.g., photometric transits), hence gravitational microlensing helps provide a more complete understanding of planet statistics by providing access to another population of planets (Bennett 2009; Gaudi 2012).

The first planetary microlensing event was discovered by Bond et al. (2004). Expected to launch in 2026, the Nancy Grace Roman Space Telescope (Roman), a NASA flagship mission, will survey ~$10^8$ stars for microlensing events (Penny et al. 2019). With less than 200 planets discovered via microlensing thus far, each new planetary microlensing analysis facilitates the calibration of theory and influences the science goals and operational plan of large-scale missions such as Roman.

A recent statistical analysis of planetary signals discovered using gravitational microlensing implied that cold, Neptune-mass planets are likely to be the most common type of planets beyond the snow line (Suzuki et al. 2016; Jung et al. 2019). This was inferred from a break in the planet-to-host-star mass-ratio function for a mass-ratio $q \sim 10^{-4}$, with the break resulting in a peak at Neptune-mass planets. Although the Suzuki et al. (2016) sample generally supports the planet distribution predictions from core accretion theory population synthesis models for planets beyond the snow line (Ida & Lin 2004; Mordasini et al. 2009), the existence of the peak in the planet-to-host-star mass-ratio distribution at Neptune-mass planets in the sample distribution presents issues. Specifically, these models of the planet distribution predict a dearth of sub-Saturn-mass planets, which conflicts with the microlensing observations (Suzuki et al. 2018). However, the Suzuki et al. (2016) sample consisted of only 30 exoplanets. This small sample size combined with the apparent contradiction emphasizes the importance of additional analyses, such as the one presented in this work. Formation models that propose a shortage of cold sub-Saturn-mass planets (Ida & Lin 2004; Mordasini et al. 2009; Ida et al. 2013) would be contested by such a population of planets (Suzuki et al. 2018; Zang et al. 2020; Terry et al. 2021), and revised formation models would be required (Ali-Dib et al. 2022).

## 2. OBSERVATIONS AND DATA REDUCTION

The microlensing event MOA-2020-BLG-208 was discovered by the Microlensing Observations in Astrophysics collaboration (MOA) and first alerted on 2020 August 11. The event was located at the J2000 equatorial coordinates (R.A., decl.) = ($17^{\rm h}\,53^{\rm m}\,43.80^{\rm s}$, $-32°35'21.52''$), and Galactic coordinates ($l$, $b$) = (357.7569650°, −3.3694423°) in the MOA-II field 'gb3'. The MOA observations were performed using the purpose-built 1.8m wide-field MOA telescope located at Mount John Observatory, New Zealand, and were taken at a 15-minute cadence with the MOA-Red filter. The MOA-Red filter corresponds to a customized wide-band similar to a sum of the Kron-Cousins $R$ and $I$ bands (600 nm to 900 nm). Additional observations were made by MOA in the visual band using the MOA-$V$ filter. The photometry in these filters was performed in real-time by the MOA pipeline (Bond et al. 2001) based on the difference imaging method of Tomaney & Crotts (1996). A re-reduction was carried out using the method detailed in Bond et al. (2017) resulting in photometry calibrated to phase-III of the Optical Gravitational Lensing Experiment (OGLE-III, Szymański et al. 2011).

Observations of the event, particularly of the primary lens event, were made by several other collaborations.

At UT 16:45 on 6 September 2020 (HJD$'$ = 9099.2), the MAP Follow-Up Collaboration and the Microlensing Follow Up Network ($\mu$FUN) found that this event could become a high magnification within two days based on the real-time MOA data and thus conducted follow-up observations. Their follow-up observations were taken using the 1.0m telescope of the Las Cumbres Observatory (LCO) global network (Brown et al. 2013) located at the South African Astronomical Observatory, South Africa (LCOS), the 0.4m telescopes at Auckland Observatory (Auck) and Possum Observatory (POS), the 0.36m telescopes at Kumeu Observatory (Kumeu), Klein Karoo Observatory (KKO), Turitea Observatory (Turitea) and Farm Cove Observatory (FCO), and the 0.3m Perth Exoplanet Survey Telescope (PEST) at Australia. The LCOS data were reduced using a custom pipeline based on the ISIS package (Alard & Lupton 1998; Alard 2000; Zang et al. 2018), and the $\mu$FUN data were reduced using DoPHOT (Schechter et al. 1993).

In addition, this event was observed by the Korea Microlensing Telescope Network (Kim et al. 2016) with two 1.6m telescopes located at the Siding Spring Observatory, Australia and the South African Astronomical Observatory, South Africa (the site at CTIO in Chile was closed due to covid).

The Observing Microlensing Events of the Galaxy Automatically project (OMEGA) observed the event with 1.0m telescopes located at the South African Astronomical Observatory, South Africa, using the Las Cumbres Observatory network of robotic telescopes (Brown et al. 2013). The OMEGA data includes SDSS-i' and SDSS-g' bands, with data reduced on a filter basis and uses the pyDanDIA photometry pipeline (pyDanDIA Contributors 2017).

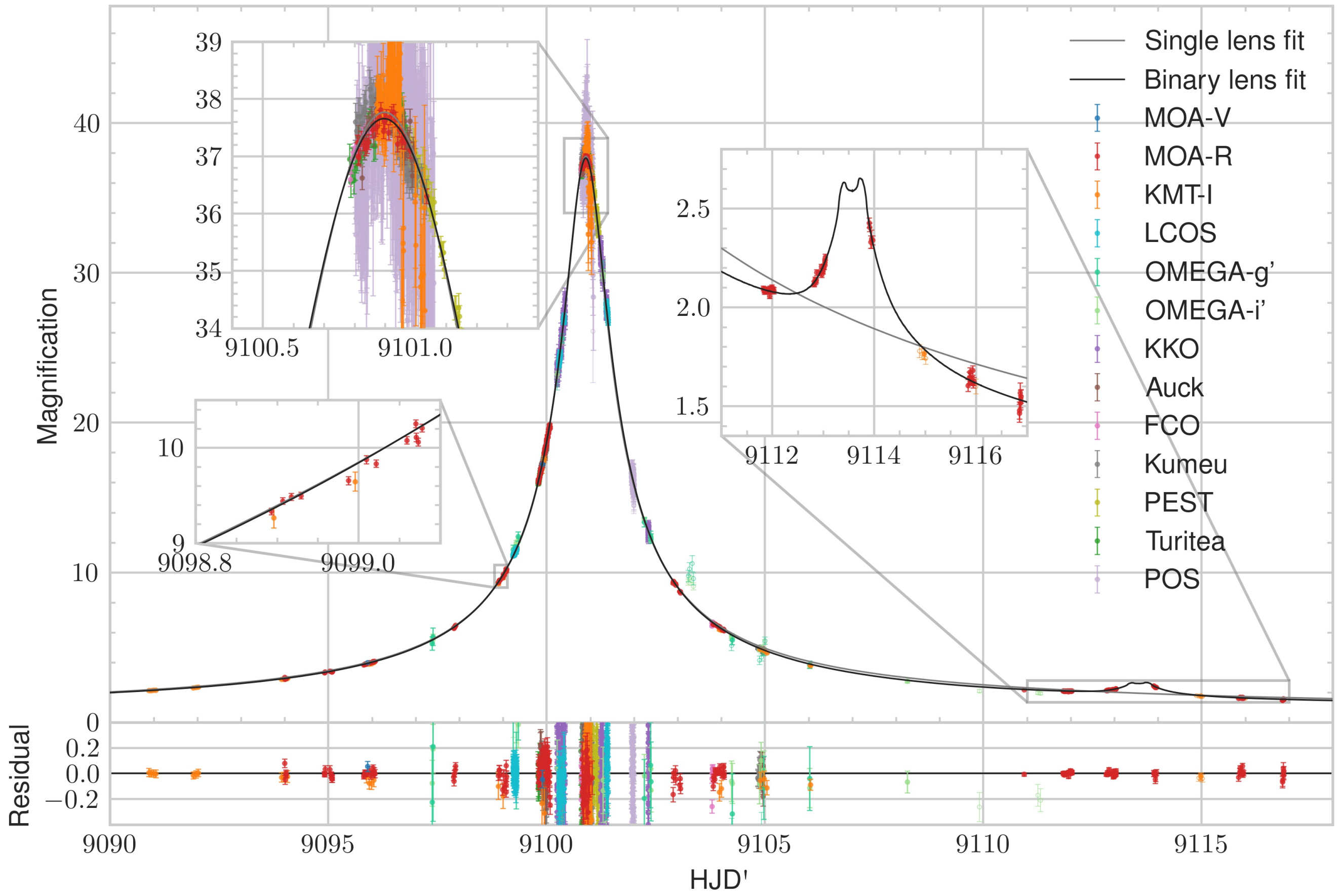

**Figure 1.** Our best-fit planetary model for a wide-orbit solution for the MOA-2020-BLG-208 event light curve, shown in black. The plot shows magnification over time (HJD'). The lower panel shows the residual of the observations from the fit model. Unfilled observation points are observations that were excluded from the fitting due to poor seeing or a $\chi^2$ cut. See text for instrument details. Zoomed inset figures show the peak of the primary magnification, the anomaly, and a portion of light curve with a significant fitting difference compared to the close-orbit solution (see Figure 2). The best-fit single-source single-lens model is shown in gray.

## 3. LIGHT CURVE MODEL

The primary lens peak of the MOA-2020-BLG-208 event light curve (see Figure 1) generally resembles a Paczýnski curve (Paczynski 1986), which is the expected shape for a microlensing event with a single lens. The deviation from the Paczýnski curve occurs in the form of a secondary anomaly near $\mathrm{HJD}' = 9113.6$[1]. The secondary anomaly in concert with the primary event is suggestive of a binary lens system comprised of a star and a companion. To model the distribution of likely properties (e.g., mass-ratio, orbital separation) that define this binary lens system, we apply the method described by Bennett (2010). To perform this modeling and analysis process, we include photometric measurements from the 15 instrument data sets that are listed in Section 2.

### 3.1. *Model fitting parameters*

The approach presented by Bennett (2010) uses an image-centered, ray-shooting method (Bennett & Rhie 1996) combined with the Metropolis-Hastings algorithm (Metropolis et al. 1953; Hastings 1970), which convergences to find $\chi^2$ minima. The parameters of our modeling consist of: the Einstein crossing time ($t_\mathrm{E}$); the time that the minimum separation of lens and source occurs ($t_0$); the minimum separation between source and lens as seen by the observer ($u_0$); the separation of the two masses of the binary lens system during the event ($s$); the counter-clockwise angle between the lens-source relative motion projected onto the sky plane and the binary lens axis ($\theta$); the mass-ratio between the secondary lens and the primary lens ($q$); the source radius crossing time ($t_*$); two values to model parallax in polar coordinates ($r_{\pi_\mathrm{E}}$ and $\theta_{\pi_\mathrm{E}}$); the source flux for each instrument $i$ ($f_{s,i}$); and the blend flux per instrument $i$ ($f_{b,i}$).

[1] HJD' = HJD − 2,450,000

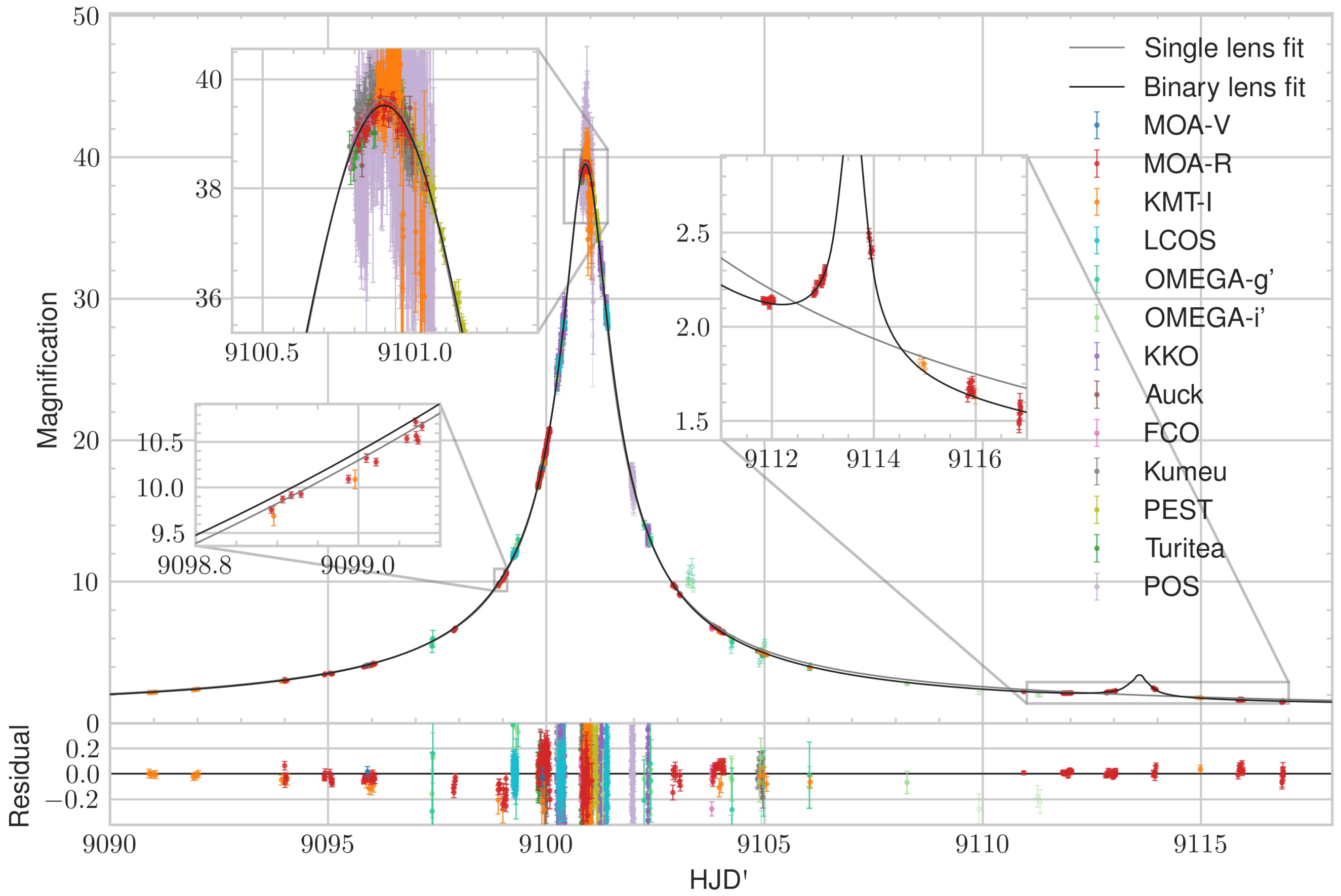

**Figure 2.** Our best-fit planetary model for a close-orbit solution for the MOA-2020-BLG-208 event light curve. All other details are the same as in Figure 1.

The parameters $t_\mathrm{E}$, $t_0$ and $u_0$ are the common parameters for the single-lens model, while $s$, $\theta$ and $q$ are the additional parameters for a binary lens system model. Both length parameters, $u_0$ and $s$, are normalized by the angular Einstein radius $\theta_\mathrm{E}$, defined by

$$\theta_\mathrm{E} = \sqrt{\frac{4GM_L}{c^2 D_S}\left(\frac{D_S}{D_L} - 1\right)}, \tag{1}$$

where $G$ is the gravitational constant, $M_L$ is the total mass of the lens system, $c$ is the speed of light, $D_S$ is the observer-source distance, and $D_L$ is the observer-lens distance. The source radius crossing time, $t_*$, is a parameter used to take into account finite source effects,

$$t_* = \rho t_\mathrm{E} = \frac{\theta_*}{\theta_\mathrm{E}} t_\mathrm{E}, \tag{2}$$

where $\rho$ is the source angular radius in Einstein units, and $\theta_*$ is the source angular radius.

The microlens parallax is denoted $\pi_\mathrm{E}$, with

$$\pi_\mathrm{E} = \frac{\pi_\mathrm{rel}}{\theta_\mathrm{E}}, \tag{3}$$

where $\pi_\mathrm{rel}$ is the lens-source relative parallax (Gould et al. 2004; Gould 1992). Our model fits the parallax parameters in polar coordinates. The equivalent Cartesian coordinates are given by $x_{\pi_E} = r_{\pi_\mathrm{E}} * \cos(\theta_{\pi_\mathrm{E}})$ and $y_{\pi_E} = r_{\pi_\mathrm{E}} * \sin(\theta_{\pi_\mathrm{E}})$.

To account for blending with nearby unlensed stars and different photometric systems, we normalize the flux data from each data source independently to minimize the $\chi^2$. As the observed brightness is linearly dependent on the blend and source fluxes, all fluxes from a single data source are normalized together to determine the normalization that produces the minimal $\chi^2$. The total flux $F_i(t)$ for time $t$ for passband $i$ is given by:

$$F_i(t) = A(t, \mathbf{x}) f_{s,i} + f_{b,i}, \tag{4}$$

where $A(t, \mathbf{x})$ is the magnification of the event at time $t$ for a set of lens model parameters $\mathbf{x} = (t_E, t_0, u_0, s, q, t_*)$, $f_{s,i}$ is the unlensed source flux for passband $i$, and $f_{b,i}$ is the blended flux for passband $i$. Each instrument's passbands are independently normalized, as the relative scales of the source and blend fluxes are dependent on the instrument and the method used for difference imaging. For example, the method presented by Bond et al. (2017) is used to process the MOA data, which normalizes the target flux to match the flux of the nearest star-like object in the reference frame.

### 3.2. *Fitting procedure*

Our fitting procedure begins with a manual rough estimate selection of $t_E$, $t_0$, $u_0$, and $t_*$. With $t_E$, $t_0$, $u_0$, and $t_*$ fixed at these selected values, we run a grid search of model fits varying $s$, $q$, and $\theta$ as described in Bennett (2010). We then select the best-fit set of parameters from this grid search for local minima (which include models from both wide-orbit and close-orbit solutions). We repeat the remaining steps for each minima model explored.

We use an MCMC fitting method (Bennett 2010) based on the Metropolis-Hastings algorithm. Our process includes an initial simulated annealing fitting, a renormalization, a second simulated annealing fitting on the renormalized values, an initial MCMC run to determine the covariance of the parameters, and finally an MCMC run with the covariance. We use the previously determined best fit grid search model parameters for $s$, $q$, and $\theta$, the estimated values for $t_E$, $t_0$, $u_0$, and $t_*$, and a value of 0 for $r_{\pi_E}$ and $\theta_{\pi_E}$ as the initial state of the MCMC algorithm. The MCMC varies all parameters during the fitting process. The acceptance of a step in the Metropolis-Hastings algorithm in our case is based on the $\chi^2$ of the fit of the model to the data. Final reported $\chi^2$ statistics use the renormalization of the overall best-fit model. This model produced the best $\chi^2$ result regardless of which renormalization was applied.

## 4. LIGHT CURVE MODELING RESULTS

In Figure 1, we show the best-fit model of our light curve fitting. This model is a single source binary lens wide-orbit model with the lens parameters shown in the "Wide model best-fit" column of Table 1. As a comparison, the best-fit single source single lens model is also shown in Figure 1. As expected, this single source single lens fit does not explain the anomalous data well.

In Table 1, we also show the best-fit single source binary lens close-orbit model for comparison. The light curve of the best-fit model for the close-orbit is shown in Figure 2. Here we find the wide-orbit model is favored over the close-solution model with a difference in $\chi^2$ of -478.8.

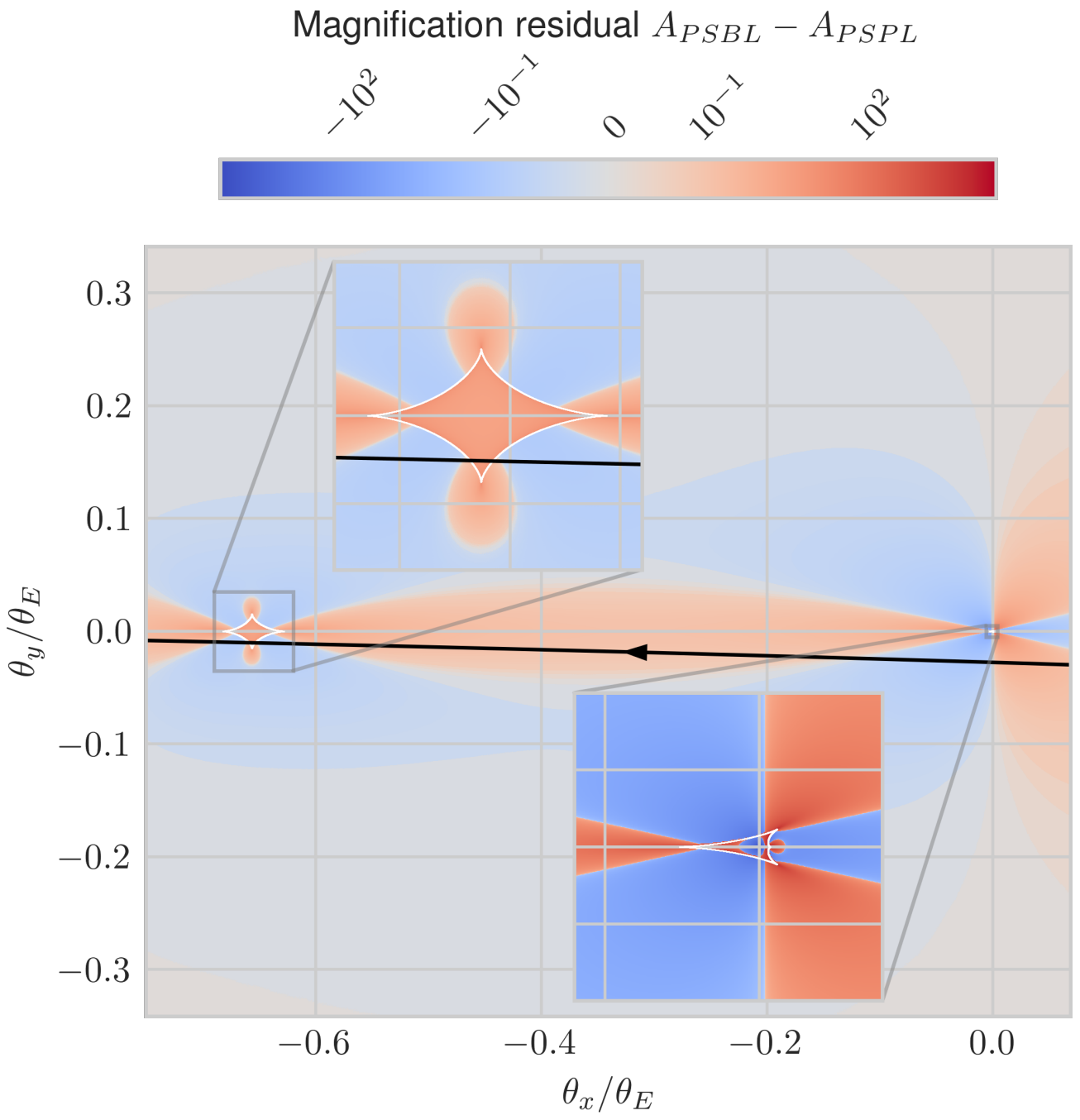

**Figure 3.** The source trajectory and magnification pattern of the wide-orbit best-fit solution. The magnification pattern shows the difference between the point-source-binary-lens (PSBL) and the point-source-point-lens (PSPL) models. White lines represent the caustic.

Figure 3 shows the source trajectory and magnification pattern of the wide-orbit best-fit solution. Figure 4 shows the equivalent for the close-orbit best-fit solution.

The distribution of MCMC states for the run with covariance can is shown in Figure 5 for the wide-orbit solution, and Figure 6 for the close-orbit solution.

Next, we obtain the dereddened color and corrected magnitude of the source star from the instrumental MOA-II magnitudes, which are calibrated via the procedure described by Bond et al. (2017). The color-magnitude diagram of the stars near MOA-2020-BLG-208 is shown in Figure 7. Using the unextincted red clump centroid as predicted by Nataf et al. (2013), we estimate the color and magnitudes shown in Table 2. Using the Boyajian et al. (2014) restricted analysis of stars with $3900 < T_{\rm eff} < 7000$, we use the color and magnitude to estimate the angular size of the source star. Using this, we estimate the Einstein radius of the system. These and other estimated source and source-lens properties are shown in Table 2.

## 5. GALACTIC MODEL ANALYSIS RESULTS

Various properties of the lens system, such as lens mass and lens distance, cannot be directly inferred with-out microlensing parallax and lens brightness measurements. These measurements do not exist for the MOA-2020-BLG-208 event, as parallax is not reliably detected and the angular source size cannot be directly measured. Indeed, the wide-solution microlensing parallax is not well constrained by our light curve fitting procedure, as is seen by the high uncertainty in the parallax parameters $r_{\pi_{\rm E}}$ and $\theta_{\pi_{\rm E}}$ in Table 1. Hence, we estimate additional lens system properties using the Bayesian analysis galactic model provided by Bennett et al. (2014). This model allows its posterior distributions to be influenced by a prior based on the host mass. Specifically, it allows the posterior distributions to rely on either a mass function that assumes that all stars have an equal planet-hosting probability or one that assumes planets are more likely to orbit around more massive stars.

| | Wide model distribution | Wide model best-fit | Close model distribution | Close model best-fit |
|---|---|---|---|---|
| $\chi^2$ | | 2768.2 | | 3247.1 |
| $t_{\rm E}$ (days) | $19.336^{+0.074}_{-0.071}$ | 19.337 | $20.226^{+0.079}_{-0.078}$ | 20.218 |
| $q$ | $3.17^{+0.28}_{-0.27} \times 10^{-4}$ | $3.10 \times 10^{-4}$ | $5.72^{+0.26}_{-0.26} \times 10^{-4}$ | $5.86 \times 10^{-4}$ |
| $t_0$ (HJD′) | $9100.9076^{+0.0010}_{-0.0010}$ | 9100.9076 | $9100.8835^{+0.0007}_{-0.0007}$ | 9100.8833 |
| $u_0$ | $0.02772^{+0.00018}_{-0.00018}$ | 0.02763 | $0.02542^{+0.00013}_{-0.00013}$ | 0.02551 |
| $t_*$ (days) | $0.298^{+0.018}_{-0.014}$ | 0.287 | $0.110^{+0.020}_{-0.056}$ | 0.130 |
| $s$ | $1.3807^{+0.0018}_{-0.0018}$ | 1.3805 | $0.73817^{+0.00091}_{-0.00092}$ | 0.73809 |
| $\theta$ (rad) | $3.112^{+0.004}_{-0.004}$ | 3.113 | $0.02837^{+0.00793}_{-0.00676}$ | 0.02804 |
| $r_{\pi_{\rm E}}$ | $0.0666^{+0.1244}_{-0.0830}$ | 0.0433 | $0.527^{+0.057}_{-0.039}$ | 0.515 |
| $\theta_{\pi_{\rm E}}$ (rad) | $2.92^{+0.38}_{-2.56}$ | 3.18 | $-1.80^{+0.44}_{-0.32}$ | $-1.88$ |

**Table 1.** A comparison of the parameters for the best-fit and median MCMC distribution models for both close and wide solutions. The MCMC distribution value also includes the upper and lower bounds of a 68.27% confidence interval.

| Property | MCMC Median |
|---|---|
| Source magnitude $I_{\rm S,0}$ | $15.008^{+0.075}_{-0.077}$ |
| Source magnitude $K_{\rm S,0}$ | $13.79^{+0.12}_{-0.17}$ |
| Source color $(V-I)_{\rm S,0}$ | $0.997^{+0.092}_{-0.100}$ |
| Source angular radius $\theta_\star$ ($\mu$as) | $4.13^{+0.45}_{-0.42}$ |
| Einstein radius $\theta_{\rm E}$ (mas) | $0.267^{+0.033}_{-0.029}$ |
| Lens-source proper motion $\mu_{\rm rel,G}$ (mas yr$^{-1}$) | $5.04^{+0.62}_{-0.55}$ |

**Table 2.** Source and lens-source system properties.

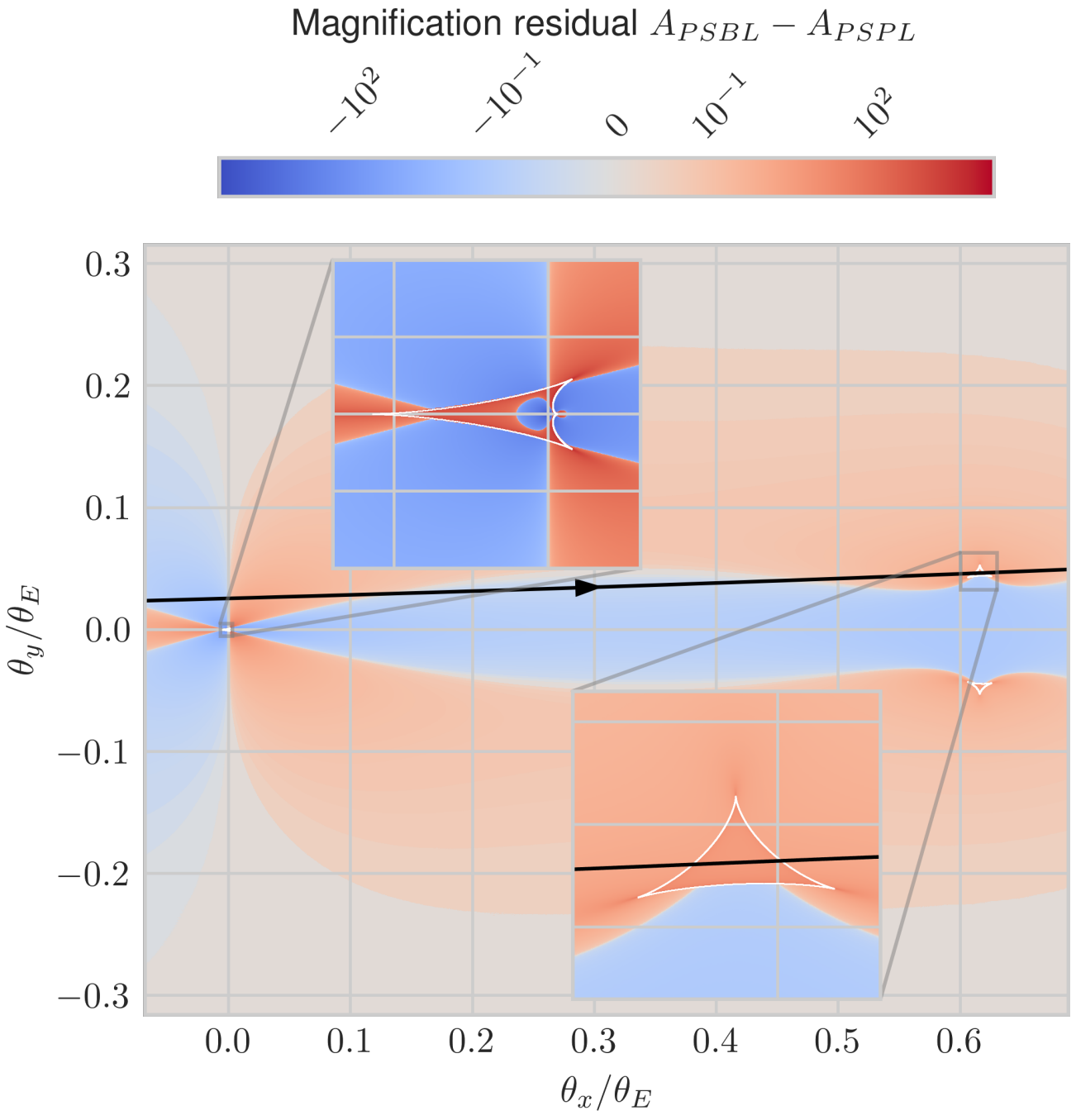

**Figure 4.** The source trajectory and magnification pattern of the close-orbit best-fit solution. The magnification pattern shows the difference between the point-source-binary-lens (PSBL) and the point-source-point-lens (PSPL) models. White lines represent the caustic.

We infer the lens properties posterior distributions using the MCMC states described in Section 3 as the input, excluding the parallax parameters. Along with the MCMC states as input, we provide the model with the estimated angular source radius and I-band extinction that we calculate as described in Section 2. To facilitate high-angular resolution follow-up observations, we additionally run the galactic model inference for the K-band, using an extinction that we calculate via the method provided by Gonzalez et al. (2012); Nishiyama et al. (2009). For MOA-2020-BLG-208, we calculate this extinction to be $A_K = 0.2062$. For both bands, we infer the lens parameters using two different input mass priors. The difference between these two priors comes in the form of varying $\alpha$ in the power-law stellar mass function defined by $P \propto M^{\alpha}$.

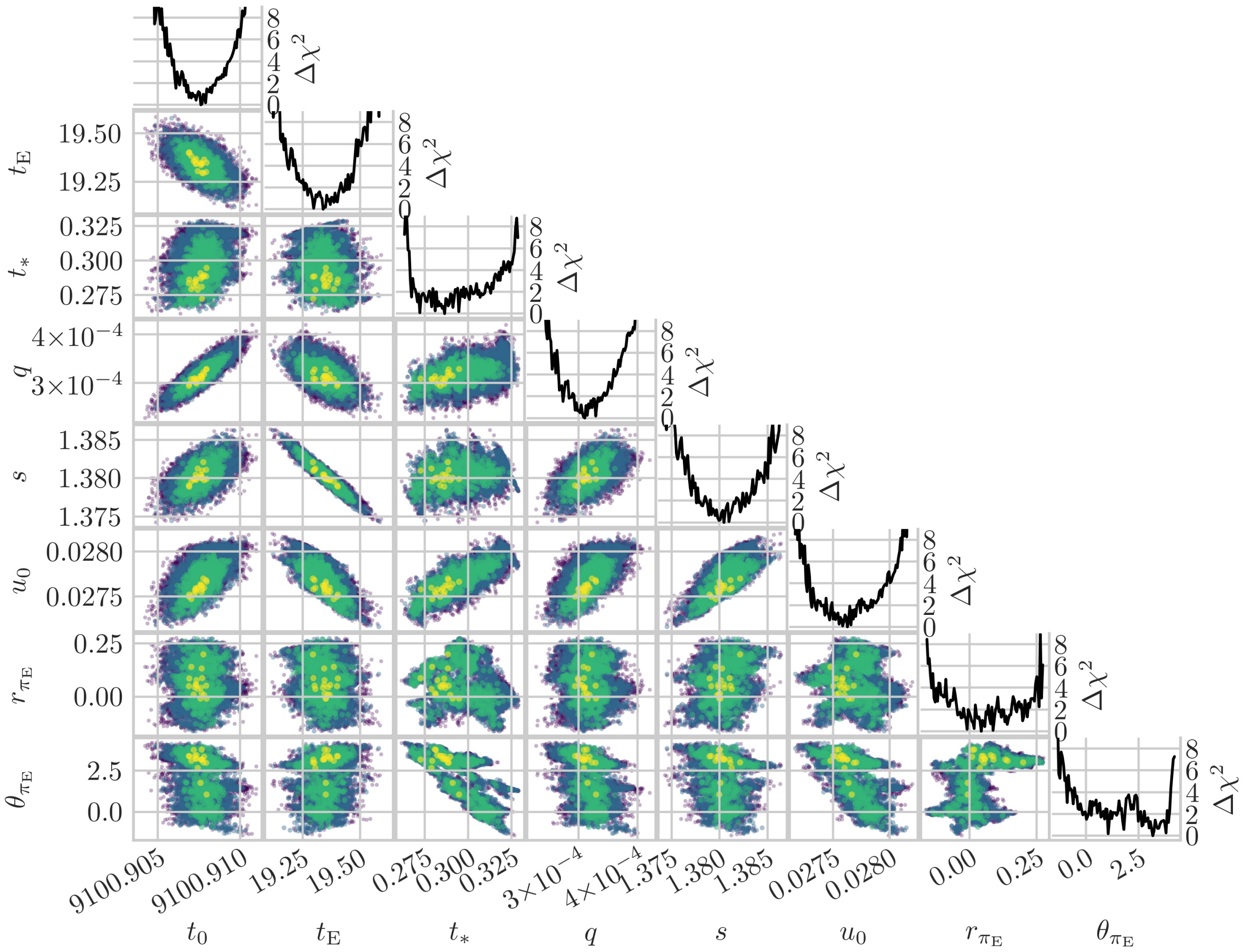

**Figure 5.** The marginalized posterior distributions of the wide model's MCMC states. The thresholds for the data point colors are for $\chi^2$ at 1, 4, 8, and 16, with samples above 16 not shown. Compare to the close-solution model in Figure 6.

| | Prior uniform in M | Prior proportional to M |
|---|---|---|
| Source distance (kpc) | $9.15^{+1.04}_{-1.16}$ | $9.02^{+1.04}_{-1.09}$ |
| I-band lens magnitude | $24.0^{+1.7}_{-2.8}$ | $22.3^{+2.3}_{-2.1}$ |
| Planet mass ($M_\oplus$) | $46^{+42}_{-24}$ | $69^{+37}_{-34}$ |
| Host mass ($M_\odot$) | $0.43^{+0.39}_{-0.23}$ | $0.66^{+0.35}_{-0.32}$ |
| Projected separation (AU) | $2.74^{+0.43}_{-0.47}$ | $2.88^{+0.40}_{-0.41}$ |
| Lens distance (kpc) | $7.49^{+0.99}_{-1.13}$ | $7.81^{+0.93}_{-0.93}$ |
| K-band lens magnitude | $20.7^{+1.3}_{-2.1}$ | $19.5^{+1.7}_{-1.7}$ |

**Table 3.** This table shows the median values of our galactic model distribution as well as the upper and lower bounds of a 68.27% confidence interval (1 $\sigma$). These results are for the wide-orbit model.

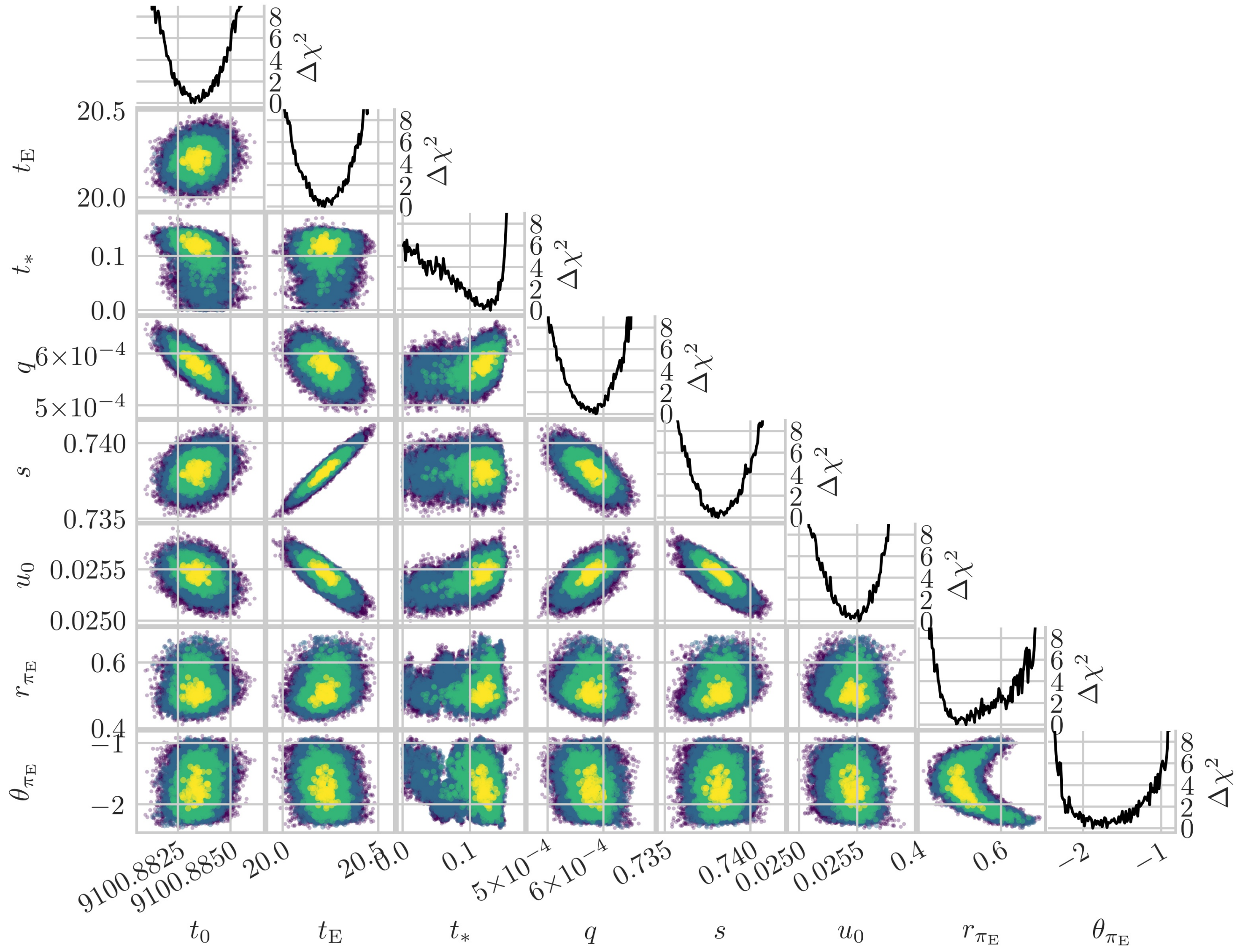

**Figure 6.** The marginalized posterior distributions of the close model's MCMC states. The thresholds for the data point colors are for $\chi^2$ at 1, 4, 8, and 16, with samples above 16 not shown. Compare to the wide-solution model in Figure 5.

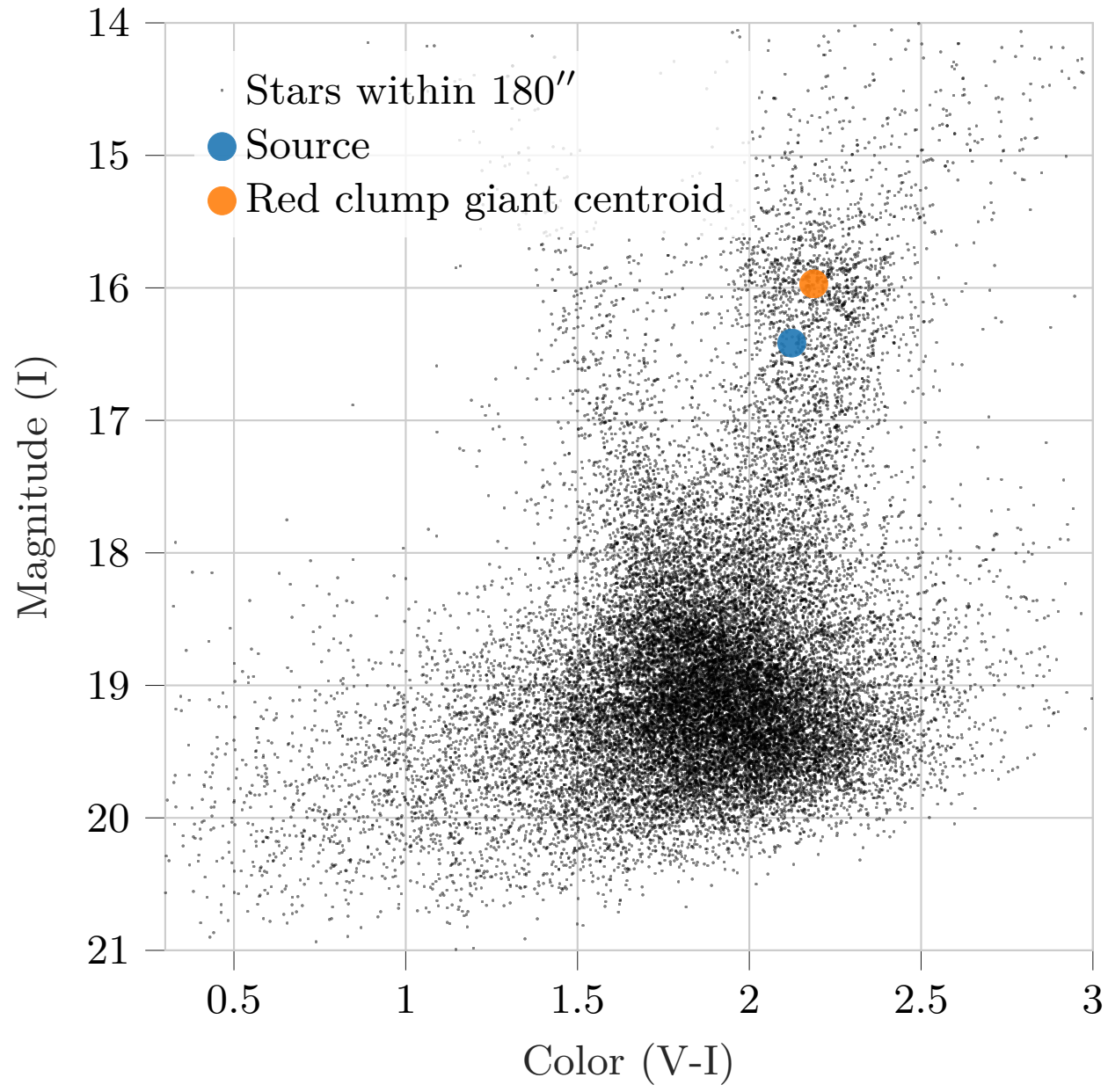

**Figure 7.** Color-magnitude diagram of the stars in the OGLE-III catalog within 180" of MOA-2020-BLG-208. The blue dot shows the source magnitude and color for the best-fit wide-orbit model (see Table 1).

In one case, we set $\alpha = 0$, which provides a mass prior that assumes all stars have an equal probability of hosting planets. This is a common prior assumption in many existing planetary microlensing analyses and related statistical population studies (e.g. Cassan et al. 2012). Figure 8 shows the Galactic model posterior distributions with $\alpha = 0$. The median values of the distributions as well as the upper and lower bounds of a 68.27% confidence interval are shown in Table 3.

However, several works have suggested that the probability of hosting a planet scales in proportion to the host star mass, including cases in radial velocity samples (Johnson et al. 2007, 2010), in direct imaging samples (Nielsen et al. 2019), and in revised analyses of microlensing samples with additional follow-up observations (Bhattacharya et al. 2021), suggesting mass prior with $\alpha = 0$ may be incorrect. Thus, following the analysis of Silva et al. (2022), we also present an analysis using the power-law stellar mass function where $\alpha = 1$, which provides a mass prior that assumes more massive stars have a greater probability of hosting planets. Figure 9 shows the Galactic model posterior distributions with $\alpha = 1$.

In this Bayesian analysis, we assume that the planet hosting probability does not depend on Galactocentric distance (Koshimoto et al. 2021).

The galactic model by Bennett et al. (2014) does not consider the possibility of a nearby source. Therefore, we performed a brief analysis using the galactic model by Koshimoto et al. (2021); Koshimoto et al. (2021); Koshimoto & Ranc (2022), which does include the possibility of nearby sources. From this, we found that probability of a source distance $D_S < 4$ kpc is less than $1.31 \times 10^{-6}$, suggesting a nearby source is unlikely.

## 6. ALTERNATIVE MODELS

### 6.1. *Evidence against a binary source model*

We also attempted to fit the observed data to a binary source model. The results of this modeling suggest the observed data does not fit well to a binary source model.

For the binary source modeling, we restricted the data sources to observatories that obtained data near the anomalous event, namely, MOA and KMT data. Each of these instruments have relatively narrow pass bands. With these data, we fit both a binary lens and binary source model. The resulting binary lens model fit was similar to the wide-solution model shown in Section 4. A comparison of these two models is shown in Table 4.

The $\chi^2$ value of the binary source model is 154.1 more than that of the binary lens model, suggesting a strong preference toward the binary lens model. Furthermore, this best fit binary source model attributes an improbable blue color to the second source (see Figure 10). Attempts to restrict the second source to a more likely color result in significant increases to the $\chi^2$ value of the model fit.

With both the significant $\chi^2$ preference toward the binary lens model and the best fit binary source model having unlikely physical parameters, we infer this event is unlikely to have been caused by a binary source.

### 6.2. *Close-orbit model orbital motion*

The parallax parameters of our close-orbit model are relatively constrained. This is likely due to the close-solution model being the incorrect model resulting in spurious fit values, as the wide-solution model presents a significantly improved $\chi^2$ value. However, the constraints on the parallax parameters may also have resulted from a degeneracy with orbital motion, which was not included in the above modeling. To test this possibility, we performed preliminary modeling of the close-solution model with orbital motion being fitted in addition to the other parameters above. During this prelimi-nary modeling, the model fit tended toward orbital periods which are unlikely according to our galactic model. Based on our orbital period distribution from our galactic modeling results, this lens parameter modeling preferred orbital periods longer than the upper bound of our 95% confidence interval at 2089 days. For practical reasons, our analysis placed an artificial limit at the 99% confidence interval, which prevented the modeling from moving toward even longer orbital periods. Notably, these preliminary fits did not improve the $\chi^2$ compared to the close-solution without orbital motion. This can be seen as further evidence that the close-solution model is unlikely to be the correct model and that the parallax constraints are likely spurious results due to it being the incorrect solution.

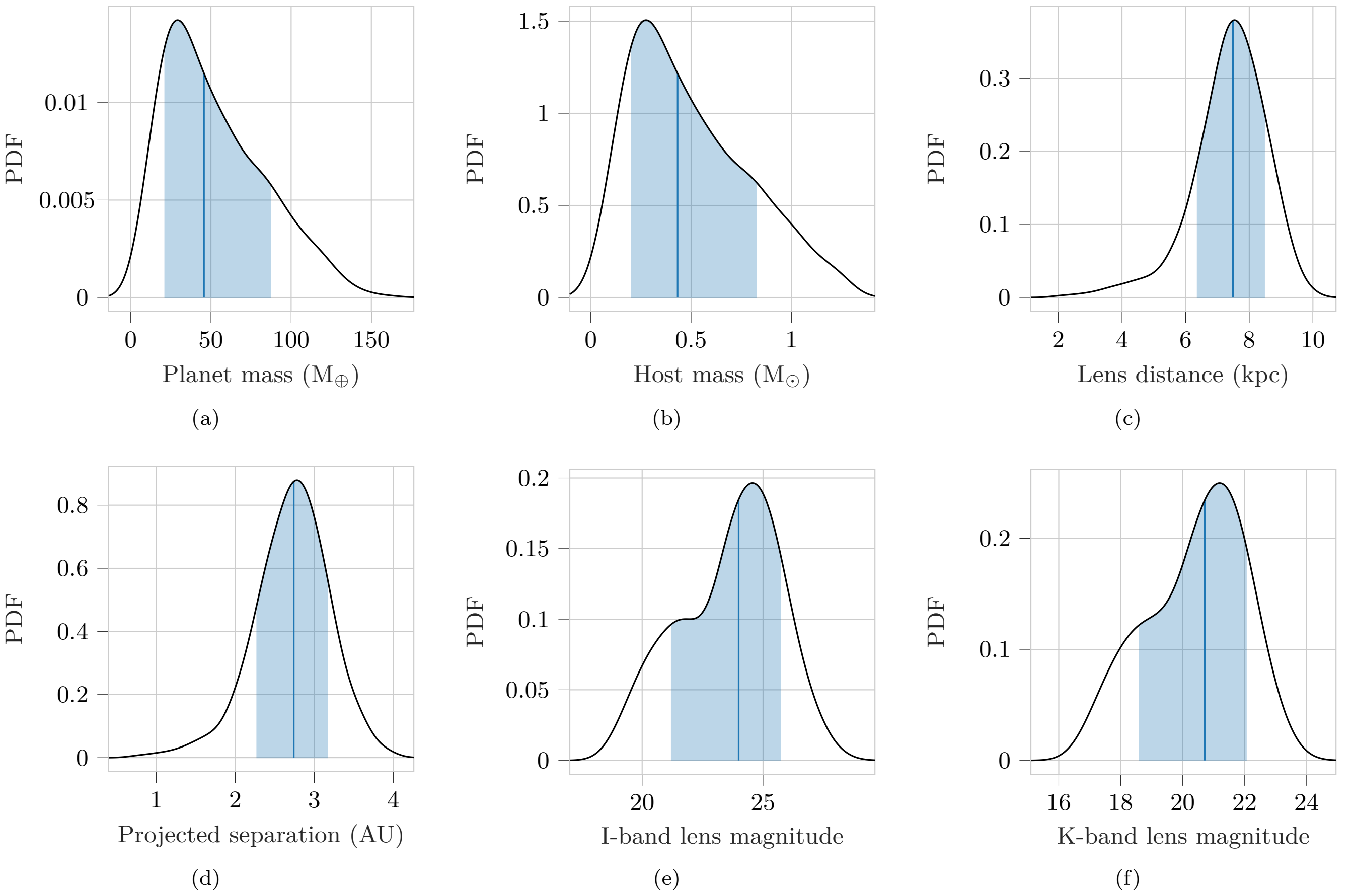

**Figure 8.** Galactic model posterior distributions with a mass prior that assumes all stars have an equal probability of hosting planets.

| | Binary source single lens model best-fit | Single source binary lens model best-fit |
|---|---|---|
| $\chi^2$ | 1579.5 | 1425.3 |
| $t_E$ (days) | 19.039 | 19.377 |
| $t_0$ (HJD′) | 9100.8982 | 9100.9058 |
| $u_0$ | 0.02708 | 0.02785 |
| $t_*$ (days) | 0.0623 | 0.324 |
| $u_{0,s_2}$ | 0.00481 | |
| $t_{0,s_2}$ | 9110 | |
| $f_{I,s_2}$ | 0.00990 | |
| $f_{R,s_2}$ | 0.0358 | |
| $q$ | | $3.91 \times 10^{-4}$ |
| $s$ | | 1.3794 |
| $\theta$ (rad) | | 3.118 |

**Table 4.** This table shows a comparison of binary-source-single-lens and single-source-binary-lens models.

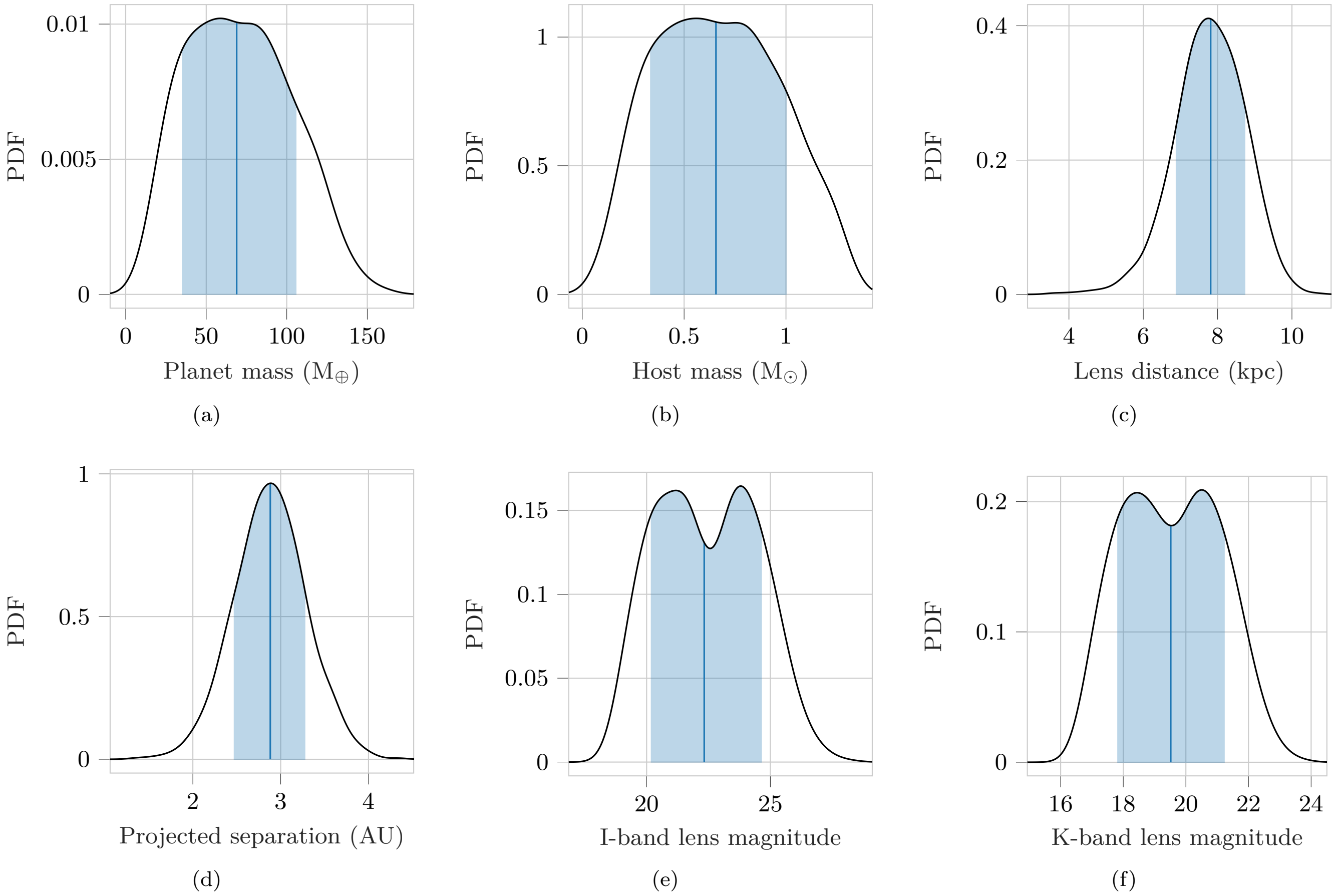

**Figure 9.** Galactic model posterior distributions with a mass prior that assumes greater star mass corresponds to a higher probability of hosting planets.

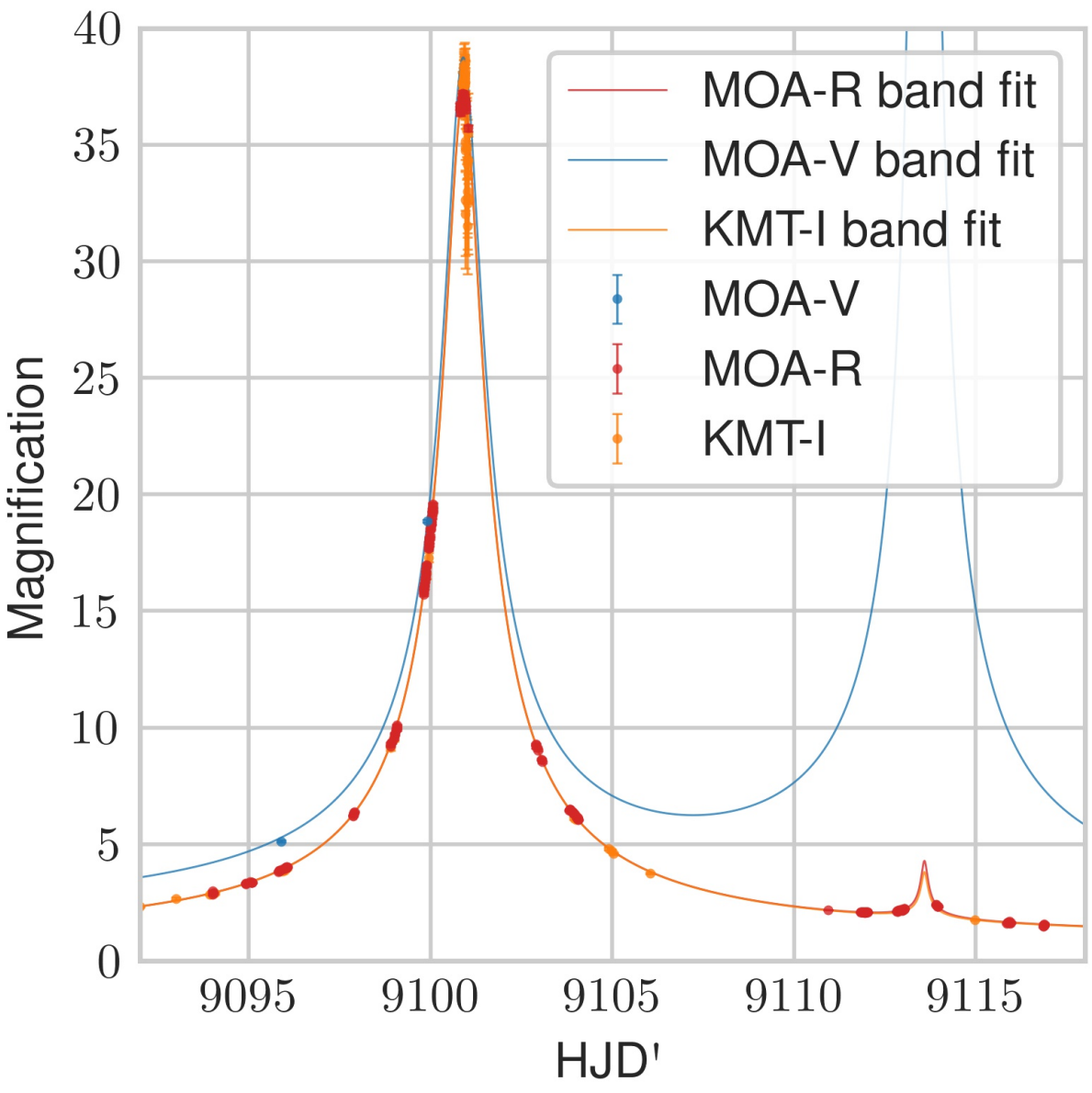

**Figure 10.** Our best-fit binary source model light curve for the MOA-2020-BLG-208 event. The extreme blueness of the second source makes this solution unlikely.

### 6.3. *Degenerate close-orbit model trajectory*

While our fitting procedure is able to obtain multiple $\chi^2$ local minima models, to ensure that it did not exclude any of importance, we explicitly fit common possible degenerate models. Notably, for the close-orbit case, a trajectory that passes on opposite side of the minor caustic compared from the best-fit model is often competitive (e.g., Han et al. 2022; Wang et al. 2022). We performed our fitting process restricting parameter space to keep the resulting model within this local minima, and found that this solution performed significantly worse than our best-fit close model, with a difference in $\chi^2$ of 354.8.

## 7. DISCUSSION

Recent exoplanet surveys have provided a plethora of planetary systems, many of which challenge established formation models (Gaudi et al. 2021). Among these surveys, gravitational microlensing surveys are apt to probe planets on significantly wider orbits (Kane 2011). Of particular note is the statistical inferences of the population of cold sub-Saturn-mass planets. Based on a limited number of detections, survey sensitivity analyses (Suzuki et al. 2016; Jung et al. 2019) suggest planets with a host-star mass-ratio, $q$, larger than $10^{-4}$ are common. Formation models that propose a shortage of cold sub-Saturn-mass planets (Ida & Lin 2004; Mordasini et al. 2009; Ida et al. 2013) would be contested by such a population of planets (Suzuki et al. 2018; Zang et al. 2020; Terry et al. 2021), and revised formation models would be required (Ali-Dib et al. 2022). Notably, Suzuki et al. (2018) presents a mass-ratio gap in models from Ida & Lin (2004); Mordasini et al. (2009) which is not identifiable in microlensing observations from Suzuki et al. (2016). Our analysis suggests a planet mass which contributes to the planet population in this gap with $q = 3.17^{+0.28}_{-0.26} \times 10^{-4}$ ($m_{\rm planet} = 69^{+37}_{-34}\ M_\oplus$ when the mass prior uses $\alpha = 1$ and $m_{\rm planet} = 46^{+42}_{-24}\ M_\oplus$ when $\alpha = 0$). Note, the planet mass is sub-Saturn even though the mass-ratio $q = 3.17^{+0.28}_{-0.26} \times 10^{-4}$ is slightly above that of Saturn. However, this still falls within of the mass-ratio gap described by Suzuki et al. (2018).

As the event was discovered via the MOA alert system, the planet model has a $\Delta\chi^2 < 100$ compared to the single lens fit, and data from other observatories support the planet model, the event qualifies for inclusion in the extended MOA-II exoplanet microlensing sample. The extended MOA-II sample is an upcoming statistical analysis of cold exoplanets detected by the MOA-II survey and is the expansion of the Suzuki et al. (2016) sample analysis. Future high-resolution angular follow-up of the lens and source may help contribute to the population of events that can be used to clarify which of the two mass priors used in this work are more appropriate.

## 8. CONCLUSION

In this work, we presented the analysis of the MOA-2020-BLG-208 gravitational microlensing event including the discovery and characterization of a new planet with an estimated mass similar to Saturn. As a cold Saturn-mass planet, this planet contributes to the evidence supporting the need for revised planetary formation models. The planet also qualifies for inclusion in the extended MOA-II exoplanet microlensing sample. We have provided evidence that the anomaly in the event is best described by a planet in a wide-solution orbit as opposed to other potential models, such as a binary source model. Our characterization was derived both from light curve modeling analysis and galactic model analysis. Notably, our galactic modeling included results from two potential mass law priors.

## 9. ACKNOWLEDGMENTS

We note that observations used in this analysis were obtained during early phases of the COVID-19 pandemic, which introduced logistical challenges for many observatories. Several collaborating observatories were unable to take measurements during this period, and those that did endured additional obstacles to do so. The MOA project is supported by JSPS KAKENHI Grant Number JSPS24253004, JSPS26247023, JSPS23340064, JSPS15H00781, JP16H06287, and JP17H02871. Work by C.R. was supported by a Research fellowship of the Alexander von Humboldt Foundation.

This research uses data obtained through the Telescope Access Program (TAP), which has been funded by the TAP member institutes. W.Zang, S.M. and W.Zhu acknowledge support by the National Science Foundation of China (Grant No. 12133005). W.Zhu acknowledges the science research grants from the China Manned Space Project with No. CMS-CSST-2021-A11.

This research has made use of the KMTNet system operated by the Korea Astronomy and Space Science Institute (KASI) and the data were obtained at the host site of SSO in Australia. J.C.Y. acknowledges support from N.S.F Grant No. AST-2108414. Y.S. acknowledges support from BSF Grant No. 2020740. Work by C.H. was supported by the grants of National Research Foundation of Korea (2020R1A4A2002885 and 2019R1A2C2085965).

E.B. gratefully acknowledge support from NASA grant 80NSSC19K0291. E.B.'s work was carried out within the framework of the ANR project COLD-WORLDS supported by the French National Agency for Research with the reference ANR-18-CE31-0002. Ł.W., K.A.R., K.K. and P.Z. acknowledge the support from the Polish National Science Centre (NCN) grants Harmonia No. 2018/30/M/ST9/00311 and Daina No. 2017/27/L/ST9/03221 as well as the European Union's Horizon 2020 research and innovation programme under grant agreement No 101004719 (OPTICON-RadioNet Pilot, ORP) and MNiSW grant DIR/WK/2018/12. Y.T. acknowledges the support of DFG priority program SPP 1992 "Exploring the Diversity of Extrasolar Planets" (TS 356/3-1).